\documentclass[conference]{IEEEtran}
\usepackage{amsmath, amssymb, amsfonts, amsthm}
\usepackage{bm}
\usepackage{cite}
\usepackage{graphicx}
\usepackage[font=footnotesize]{subcaption}
\usepackage{xcolor}
\definecolor{niceMagenta}{RGB}{199,21,133}
\usepackage{acronym}
\usepackage{stfloats}
\usepackage{enumitem}

\acrodef{OFDM}{orthogonal frequency division multiplexing}
\acrodef{AFDM}{affine frequency division multiplexing}
\acrodef{DAFT}{discrete affine Fourier transform}
\acrodef{IDAFT}{inverse discrete affine Fourier transform}
\acrodef{DFT}{discrete Fourier transform}
\acrodef{IFFT}{inverse fast Fourier transform}
\acrodef{FFT}{fast Fourier transform}
\acrodef{ISAC}{integrated sensing and communications}
\acrodef{LTV}{linear time-variant}
\acrodef{LTI}{linear time-invariant}
\acrodef{ICI}{inter-carrier interference}
\acrodef{CP}{cyclic prefix}
\acrodef{CPP}{chirp-periodic prefix}
\acrodef{CPAFDM}[CP-AFDM]{chirp-permuted AFDM}
\acrodef{OCDM}{orthogonal chirp division multiplexing}
\acrodef{OTFS}{orthogonal time frequency space}
\acrodef{V2X}{vehicle-to-everything}
\acrodef{NTN}{non-terrestrial network}
\acrodef{SAGIN}{space-air-ground integrated network}
\acrodef{UAV}{unmanned aerial vehicle}
\acrodef{URLLC}{ultra-reliable low-latency communications}
\acrodef{BER}{bit error rate}
\acrodef{SNR}{signal-to-noise ratio}
\acrodef{AF}{ambiguity function}
\acrodef{LTE}{Long-Term Evolution}
\acrodef{NR}{New Radio}
\acrodef{ADSL}{asymmetric digital subscriber line}
\acrodef{VDSL}{very-high-bit-rate digital subscriber line}
\acrodef{AFT}{affine Fourier transform}
\acrodef{LCT}{linear canonical transform}
\acrodef{FT}{Fourier transform}
\acrodef{PAPR}{peak-to-average power ratio}
\acrodef{MMSE}{minimum mean-square error}
\acrodef{CDMA}{code-division multiple access}
\acrodef{IDFnT}{inverse discrete Fresnel transform}
\acrodef{IDFT}{inverse discrete Fourier transform}
\acrodef{FLOP}{floating point operation}
\acrodef{AWGN}{additive white Gaussian noise}
\acrodef{IoT}{Internet-of-Things}

\begin{document}

\title{AFDM as a Software Upgrade of OFDM: \\ One Firmware Patch, a New Frontier}

\author{%
\IEEEauthorblockN{Hyeon Seok Rou and Giuseppe Thadeu Freitas de Abreu}
\IEEEauthorblockA{School of Computer Science and Engineering,
Constructor University Bremen, Germany \\
Emails: \{hsrou@ieee.org, gabreu@constructor.university\}}
\vspace{0.5ex}
}

\maketitle

\begin{abstract}
In this article, we start by highlighting that \ac{AFDM} can be implemented at marginal cost as a \textit{software upgrade} (firmware patch) of conventional \ac{OFDM}, enabling its adoption across the vast installed base of \ac{OFDM} infrastructure without new hardware.
This upgrade delivers four compounding benefits: 1) robustness in doubly dispersive channels for high-mobility 6G use-cases including urban vehicular communications, \ac{NTN}, and drone networks, 2) full uncoded diversity even in static \ac{LTI} channels, a less highlighted strength of \ac{AFDM}, 3) a broad secondary design space -- including \ac{PAPR} control, physical-layer security, and index modulation -- enabled by the free chirp parameters, and of course 4) native \ac{ISAC} compatibility enabling joint communication and sensing within a single waveform, a critical capability for 6G and IEEE 802.11bf.
The \ac{ISAC} capability, in particular, is linked to the full delay-Doppler path separability of the \ac{AFDM} effective channel, which establishes a structural equivalence between channel estimation and radar parameter estimation.
We also provide step-by-step Fourier-analytic derivations of the effective channel to expose these mechanisms intuitively against the \ac{OFDM}, and find that \ac{AFDM} is a unique practical path towards \ac{ISAC}-enabled multicarrier waveforms.
\end{abstract}

\acresetall

\vspace{2ex}

\section{Introduction and Motivation}

It is no exaggeration that \ac{OFDM} is the most dominant multicarrier waveform in modern communications, owing to its elegance, efficiency, and effectiveness.
In \ac{OFDM}, modulation and demodulation via sinusoidal subcarriers reduce to an \ac{IFFT}/\ac{FFT} pair, making implementation hardware-efficient and mathematically transparent -- driving adoption across 4G/5G cellular (\acs{LTE}/\acs{NR}), IEEE 802.11 Wi-Fi, DVB-T/T2 broadcasting, and \acs{ADSL}/\acs{VDSL} wireline broadband, with over five million 5G base stations deployed worldwide as of 2025.

However, the channel conditions demanded by 6G applications expose fundamental limitations.
Dense high-mobility scenarios including \ac{V2X}, \acp{NTN}, high-speed rail, and \ac{UAV} communications, in addition to higher frequency bands, inevitably result in \emph{doubly dispersive} channels with simultaneous multipath delay and Doppler frequency spread.
In such environments, Doppler shifts destroy subcarrier orthogonality of \ac{OFDM}, causing \ac{ICI} and severe performance degradation.
Furthermore, 6G and IEEE 802.11bf are expected to natively support \ac{ISAC}~\cite{LiuJSAC2022,ZhangJSTSP2021,Rou2024SPMag}, delivering communication and sensing within a single waveform.
While \ac{OFDM} has been explored for radar sensing~\cite{SturmProcIEEE2011}, a structural limitation of its effective channel prevents reliable multi-target identification.

Therefore, a change at the physical layer might be envisioned -- however, deploying a fundamentally different waveform requiring new hardware would demand retrofitting or replacing millions of base stations, which is economically and logistically impractical.

\subsection*{AFDM: Sheared Spreading of the OFDM Subcarriers}

\Ac{AFDM}, first proposed in~\cite{Bemani2023TWC}, modulates information symbols onto mutually orthogonal linear chirps, which are inherently robust to path delay and Doppler shifts -- a property long exploited in radar and spread-spectrum communications.

\begin{figure}[b]
\vspace{-5ex}
\centering
\hfil
\begin{subfigure}[b]{0.43\columnwidth}
\centering
\includegraphics[width=\linewidth]{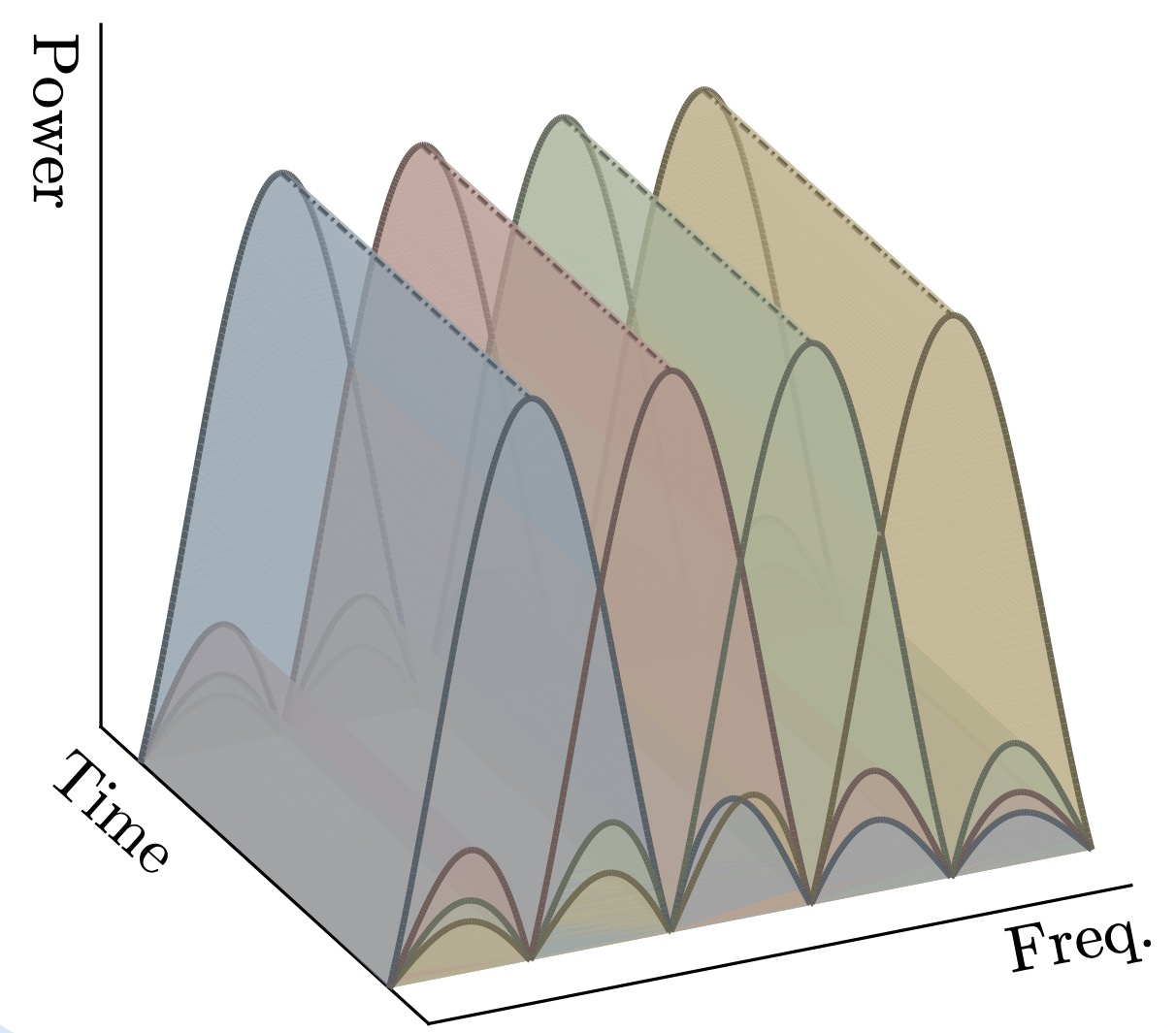}
\caption{Sinusoidal subcarriers}
\label{fig:ofdm_sub}
\end{subfigure}%
\hfil
\begin{subfigure}[b]{0.56\columnwidth}
\centering
\includegraphics[width=\linewidth]{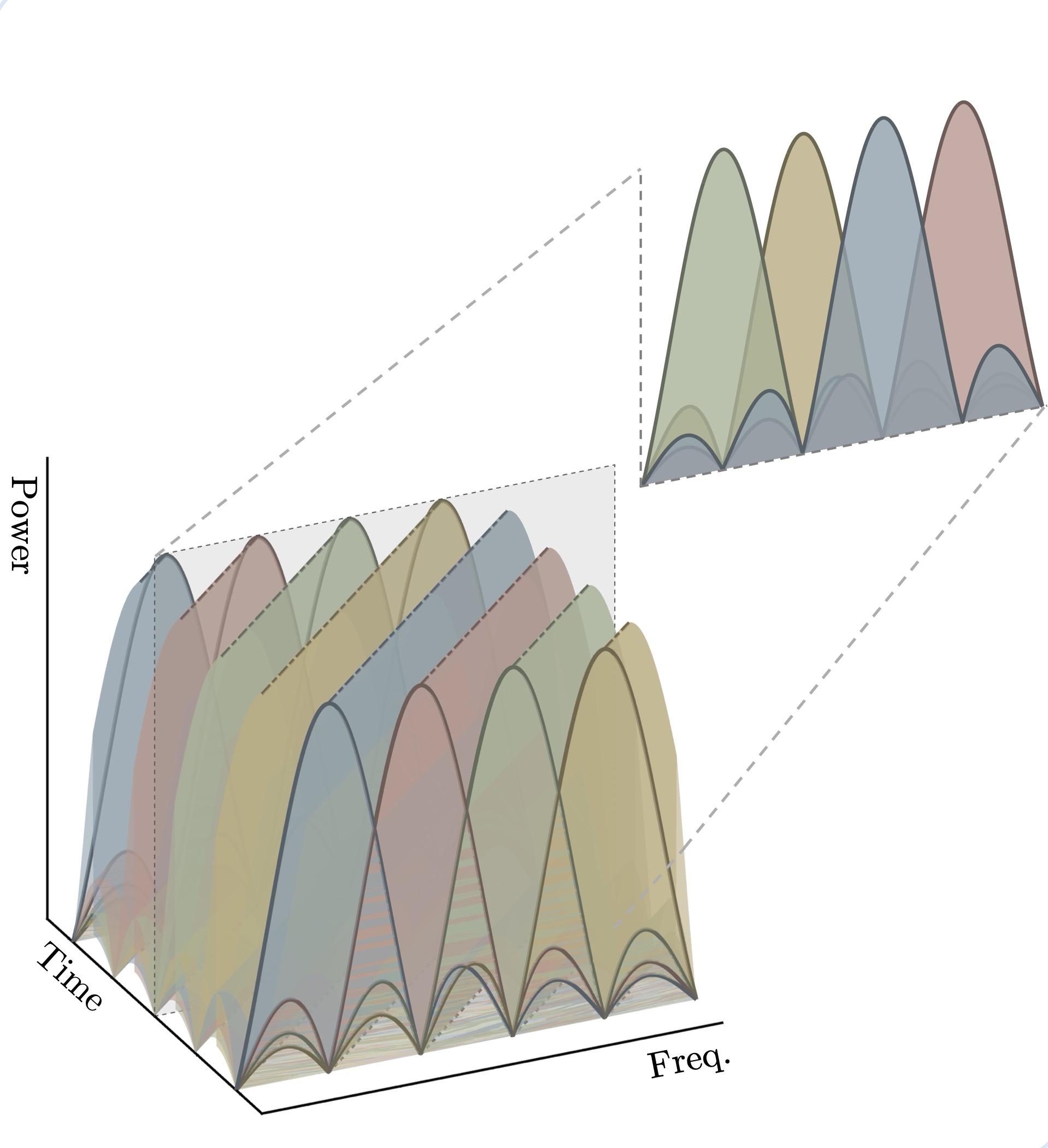}
\caption{Chirp subcarriers}
\label{fig:afdm_sub}
\end{subfigure}
\caption{Comparison of the subcarriers of OFDM and AFDM over the time-frequency plane (\textit{figure excerpt from}~\cite[Fig.~2]{Rou2026Spreading})}
\label{fig:subcarriers}
\hfil
\vspace{-2.7ex}
\end{figure}

As illustrated in Fig.~\ref{fig:subcarriers}, each \ac{AFDM} chirp subcarrier sweeps linearly across the bandwidth over the symbol duration with a freely parametrizable chirp rate, unlike the fixed-frequency sinusoidal subcarriers of \ac{OFDM}.
In other words, \ac{AFDM} subcarriers are \ac{OFDM} sinusoids subjected to a \emph{sheared spreading} in the time-frequency plane, sweeping continuously to provide frequency diversity while retaining subcarrier-wise multiplexing~\cite{Rou2026Spreading}.

\begin{figure*}[b]
    \vspace{-2ex}
\centering
\includegraphics[width=0.95\textwidth]{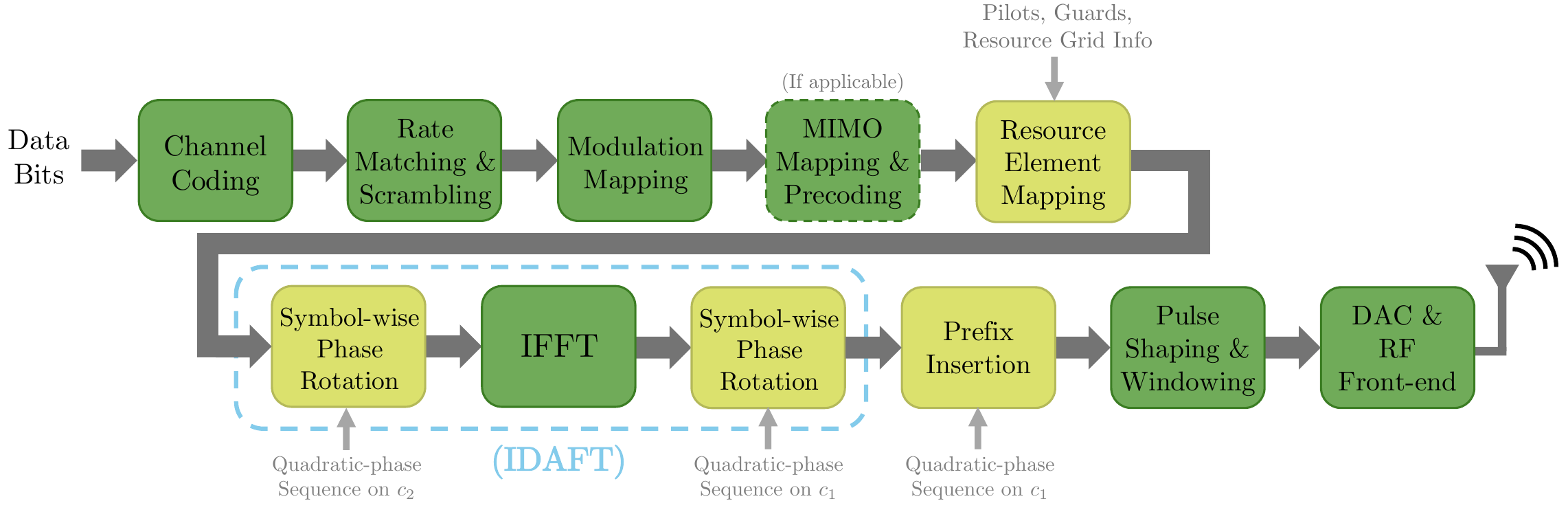}
\vspace{-1ex}
\caption{An \acs{AFDM} trasnmitter architecture illustrating the unchanged blocks from \ac{OFDM} (dark green) and blocks requiring a firmware update but keeping the hardware completely (light green) (\textit{figure excerpt from}~\cite[Fig.~4]{Rou2026Spreading}).}
\label{fig:blockdiag}
\end{figure*}

\subsection*{Paper Contributions}

A growing body of literature covers \ac{AFDM}'s analysis, channel estimation, equalization, \ac{ISAC}, and system design~\cite{Rou2025CommStd,Rou2026OpenJ,Bemani2024WCL,Ni2025TWC}, and this paper does not replicate those results but assembles the key structural insights into two central contributions:

\begin{enumerate}
\item \ac{AFDM} can be implemented directly over standard \ac{OFDM} infrastructure as a lightweight \textit{firmware patch},
\item This patch constitutes a \textit{substantial upgrade} -- enabling not only Doppler robustness and native \ac{ISAC} compatibility, but also a broad secondary waveform design space and full uncoded diversity in static \ac{LTI} channels.
\end{enumerate}

\section{What is \acs{AFDM}?}
\label{sec:what}

\Ac{AFDM} generates and multiplexes symbols onto orthogonal chirp signals via the \ac{AFT}, a four-parameter generalization of the \ac{FT} that, under certain parameter constraints, reduces to a two-parameter form with tunable chirp parameters $c_1, c_2 \in \mathbb{R}$.

Discretely, this is described as a block of $N$ information symbols $\mathbf{x} = [x_0, x_1, \ldots, x_{N-1}]^\mathsf{T} \in \mathbb{C}^{N \times 1}$ modulated onto a time-domain sequence $\mathbf{s} = [s_0, s_1, \ldots, s_{N-1}]^\mathsf{T} \in \mathbb{C}^{N \times 1}$ via the
\ac{IDAFT} as
\begin{equation}
s_n = \frac{1}{\sqrt{N}} \sum_{m=0}^{N-1} \underbrace{e^{j2\pi\left(c_2 m^2 + \frac{nm}{N} + c_1 n^2\right)}}_{\triangleq \,\phi_{n,m}(c_1, c_2)} \cdot \, x_m
\end{equation}
for $n \in \{0, 1, \ldots, N-1\}$.

Inspection of the \ac{IDAFT} kernel $\phi_{n,m}(c_1,c_2)$ shows that $c_1 = c_2 = 0$ recovers the \ac{IDFT} of \ac{OFDM}, and $c_1 = c_2 = \frac{1}{2N}$ recovers the \ac{IDFnT} of \ac{OCDM}. \Ac{AFDM} is thus a strict superset of both, with the chirp parameters as a continuous waveform-design knob.

\subsection*{\acs{AFDM} as a Software Upgrade of \acs{OFDM}}

This discrete counterpart of the \ac{AFT} admits a compact decomposition into an \ac{FFT} sandwiched between two diagonal chirp phase matrices, as 

$~$\vspace{-4ex}

\begin{equation}
\mathbf{A}_N = \mathrm{diag}(\boldsymbol{\lambda}_{c_2}) \cdot \mathbf{F}_N \cdot \mathrm{diag}(\boldsymbol{\lambda}_{c_1}) \in \mathbb{C}^{N \times N},
\label{eq:idaft_intro}
\end{equation}
where $\mathbf{A}_N$ is the $N$-point forward \ac{DAFT} matrix, $\mathbf{F}_N$ is the $N$-point forward \ac{DFT} matrix, and $\boldsymbol{\lambda}_{c_1}, \boldsymbol{\lambda}_{c_2}$ are the chirp phase sequences given by
\begin{equation}
\boldsymbol{\lambda}_{c_i} \!=\! \big[1,\, e^{-j2\pi c_i (1)^2},\, \ldots,\, e^{-j2\pi c_i (N-1)^2}\big]^\mathsf{T} \in \mathbb{C}^{N \times 1}, \!\!
\end{equation}
with $c_1, c_2 \in \mathbb{R}$ being the chirp-rate parameters of the quadratic-phase sequences.

Due to the unitarity of the two diagonal chirp matrices and the \ac{DFT} matrix, the \ac{IDAFT} is also unitary, given by
\begin{equation}
\mathbf{A}_N^{-1} = \mathbf{A}_N^\mathsf{H} = \mathrm{diag}(\boldsymbol{\lambda}_{c_1}^\mathsf{*}) \cdot \mathbf{F}_N^\mathsf{H} \cdot \mathrm{diag}(\boldsymbol{\lambda}_{c_2}^\mathsf{*}) \in \mathbb{C}^{N \times N},
\label{eq:idaft_inv}
\end{equation}
where $\boldsymbol{\lambda}_{c_1}^\mathsf{*}$ and $\boldsymbol{\lambda}_{c_2}^\mathsf{*}$ are the complex conjugates of the chirp phase sequences (negative phase progression).

As multiplication by a diagonal matrix is equivalent to an element-wise multiplication, the modulation of a vector of information symbols $\mathbf{x} \in \mathbb{C}^{N \times 1}$ via the \ac{IDAFT} at the transmitter can be rewritten as a sequence of element-wise product, an IDFT, and another element-wise product, i.e.,
\begin{equation}
\mathbf{s} = \mathbf{A}_N^\mathsf{H} \cdot \mathbf{x} = \boldsymbol{\lambda}_{c_1}^\mathsf{*} \odot \left(\mathbf{F}_N^\mathsf{H} \cdot (\boldsymbol{\lambda}_{c_2}^\mathsf{*} \odot \mathbf{x})\right) \in \mathbb{C}^{N \times 1},
\end{equation}
where $\odot$ denotes the element-wise (Hadamard) product.

The \ac{AFDM} modulator (\ac{IDAFT}) therefore requires exactly two length-$N$ element-wise phase rotations on top of the \ac{IFFT}, and symmetrically at the demodulator.

\subsubsection{AFDM modulator is a firmware update} This element-wise phase-rotation mechanism already exists in conventional \ac{OFDM} implementations and is widely adopted in practice: selected mapping for \ac{PAPR} reduction, cyclic delay diversity in \ac{LTE}, pilot phase randomization in IEEE 802.11 WLANs, and pilot scrambling in DVB-T.
Implementing \ac{AFDM} therefore requires only a firmware reconfiguration of that existing mechanism to produce the specific chirp phase sequences of the \ac{IDAFT}. 
No new hardware, no new baseband chips, as illustrated in Fig.~\ref{fig:blockdiag}, which has also been confirmed experimentally on software-defined radio platforms.

\subsubsection{Computational overhead} The $2N$ additional complex multiplications require $12N$ \acp{FLOP} on top of the $5N\log_2N$ \acp{FLOP} of the \ac{IFFT}/\ac{FFT}, adding a relative overhead of $\frac{12}{5\log_2 N}$ that decreases with $N$ and is a worthwhile trade-off for the gains described in the following.

\subsection*{Chirp-Periodic Prefix (CPP) for AFDM}

\begin{figure}[b]
\vspace{-1ex}
\centering
\includegraphics[width=0.77\columnwidth]{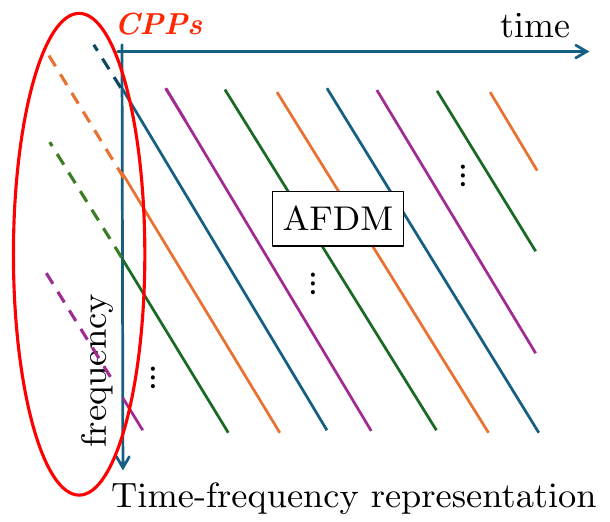}
\caption{Illustration of the required phase rotations for the \ac{CPP}.}
\label{fig:CPP}
\end{figure}

\Ac{AFDM} also requires a \ac{CPP} in place of the \ac{CP} of \ac{OFDM}.
Like the \ac{CP}, the \ac{CPP} prepends the last $N_\mathrm{CPP}$ samples ($N_\mathrm{CPP} \geq \ell_{\mathrm{max}}$) to the front of the block, where $\ell_\mathrm{max}$ is the maximum channel delay spread in samples. The only difference is that each prepended sample is additionally phase-rotated to satisfy the chirp-periodicity condition of \ac{AFDM}, as illustrated in Fig.~\ref{fig:CPP},
\begin{equation}
s[n'] = s[N+n'] \cdot e^{-j2\pi c_1 (N^2 + 2Nn')},
\label{eq:CPP}
\end{equation}
for $n' \in \{-N_\mathrm{CPP}, \ldots, -1\}$, or alternatively
\begin{equation}
\mathbf{s}_\mathrm{CPP} = \boldsymbol{\lambda}_{\mathrm{CPP}} \odot \mathbf{s}_\mathrm{CP} \in \mathbb{C}^{N_\mathrm{CPP} \times 1},
\end{equation}
where $\mathbf{s}_\mathrm{CPP} \in \mathbb{C}^{N_\mathrm{CPP} \times 1}$ is the \ac{CPP} prefix, $\mathbf{s}_\mathrm{CP} \in \mathbb{C}^{N_\mathrm{CPP} \times 1}$ is the conventional \ac{CP} that would be used for \ac{OFDM}, and $\boldsymbol{\lambda}_{\mathrm{CPP}} \triangleq [e^{-j2\pi c_1 (N^2 - 2NN_\mathrm{CPP})}, \ldots, e^{-j2\pi c_1 (N^2 - 2N)}]^{\mathsf{T}} \in \mathbb{C}^{N_\mathrm{CPP} \times 1}$ is the phase rotation sequence determined by $c_1$.

The \ac{CPP} can therefore be implemented via a firmware update, reusing the same hardware block that prepends the \ac{CP}, with only the phase-rotation coefficients updated.
Furthermore, a useful special case exists when $2Nc_1$ is an integer and $N$ is even: the phase rotations reduce to identity and the \ac{CPP} becomes the conventional \ac{CP} -- i.e., no overhead at all.

\subsection*{Resource Grid and Compatibility with OFDM}

\Ac{AFDM} operates over the same time-frequency resource grid as \ac{OFDM}. Subcarrier spacing, symbol duration, and resource element structure are all preserved, and the scheduling numerology carries over directly to the affine-frequency domain.
Dedicated \ac{AFDM} channel estimation uses a single pilot symbol with one-dimensional guard samples, recovering doubly dispersive channels with fewer pilot resources than \ac{OFDM}~\cite{Bemani2023TWC}, and the same pilot simultaneously extracts the environment delay-Doppler map, enabling \ac{ISAC} without additional sensing overhead~\cite{Bemani2024WCL,Ni2025TWC}. Conventional \ac{OFDM} pilot structures remain compatible under frequency-domain processing, facilitating gradual migration.

\subsection*{Receiver Counterpart and Key Structural Change}

The \ac{AFDM} demodulator is the conjugate counterpart of the modulator, i.e., de-rotate by $\boldsymbol{\lambda}_{c_1}$, \ac{FFT}, de-rotate by $\boldsymbol{\lambda}_{c_2}$, completing the \ac{DAFT} $\mathbf{A}_N$ from \eqref{eq:idaft_intro}.
Trivially, this is achieve also via firmware reconfiguration only.

Therefore, the principal structural difference from \ac{OFDM} lies not in the modulator or demodulator, but in channel equalization.
In \ac{OFDM}, the diagonal effective channel admits one-tap equalization.
However, in \ac{AFDM}, the banded non-diagonal effective channel requires more sophisticated receivers -- the main caveat relative to \ac{OFDM} -- though this is being actively addressed by message-passing detectors and advanced low-complexity equalizers, and remains entirely within digital baseband processing~\cite{Bemani2022ICASSP,Rou2026OpenJ}.

\vspace{3ex}
\section{Effective Channel Analysis: \\ Diversity and ISAC Mechanisms}
\label{sec:known}

This upgrade to \ac{AFDM} delivers two key gains over \ac{OFDM}, well established in the literature:
\begin{itemize}
\item Full multipath diversity order in doubly dispersive, or \ac{LTV} channels, equal to the number of unique scattering paths in the channel.
\item Inherent \ac{ISAC} compatibility arising from the delay-Doppler path separability of the effective channel, yielding a deterministic relation between the channel structure and the environment scatterers.
This results in the channel estimation--radar parameter estimation equivalence.
\end{itemize}

In what follows, we make explicit the step-by-step Fourier-analytic derivation of the effective channel -- a derivation often hidden within the broader analysis of the literature -- to expose the underlying mechanism behind full diversity and \ac{ISAC} compatibility with maximal clarity.

\subsection{The Doubly-Dispersive Channel}
\label{sec:chan}

A doubly dispersive (\ac{LTV}) channel can be modeled with $P$ propagation paths, each with complex gain $h_p$, continuous delay $\tau_p$ (in seconds), and continuous Doppler shift $\nu_p$ (in Hz).

The time-variant impulse response function is given by
\begin{equation}
h(t,\tau) = \sum_{p=1}^{P} h_p \, e^{j2\pi \nu_p t} \, \delta(\tau - \tau_p),
\label{eq:tvirf}
\end{equation}
and the corresponding received signal is given by
\begin{align}
r(t) &= \int h(t,\tau)\, s(t-\tau)\,\mathrm{d}\tau + n(t) \nonumber \\
&= \sum_{p=1}^{P} h_p \, e^{j2\pi \nu_p t} \, s(t - \tau_p) + n(t),
\label{eq:rcont}
\end{align}
where $r(t)$, $s(t)$, and $n(t)$ are respectively the time-domain continuous received, transmitted, and \ac{AWGN} signals.

Sampling at rate $f_s = \frac{1}{T_s}$ yields the discrete received signal
\begin{equation}
r[n] = \sum_{p=1}^{P} h_p \, e^{j2\pi \nu_p n T_s}
\sum_{\ell} s[n-\ell]\,\mathrm{sinc}\!\left(\ell - \frac{\tau_p}{T_s}\right) + n[n],
\label{eq:rdisc}
\end{equation}
where $r[n]$, $s[n]$, and $n[n]$ are respectively the $n$-th samples of the received, transmit, and \ac{AWGN} signal sequences with $n \in \{0,\ldots,N-1\}$, and the sinc function arises from ideal bandlimited interpolation at non-integer delay offsets\footnote{%
If practical pulse shaping is used, this leads to other interpolation functions, rigorously analyzed in~\cite{Mirabella2026JSAC,Rou2026OpenJ}.
Fractional-delay \ac{LTI} channels effectively behave as \ac{LTV} channels, and \ac{LTV} channels as denser \ac{LTV} channels.}.

For convenience, we define the normalized delay and Doppler as the digital indices (in samples),
\begin{equation}
\ell_p \triangleq \frac{\tau_p}{T_s}, \qquad f_p \triangleq N \nu_p T_s,
\label{eq:norm}
\end{equation}
and when $\ell_p \in \mathbb{Z}_{\geq 0}$, the sinc collapses to the unit delta function $\delta[\ell - \ell_p]$.

Then, with a cyclic prefix of length $N_\mathrm{CPP} \geq \ell_{\max}$ ensuring circular convolution, we arrive at
\begin{equation}
r[n] = \sum_{p=1}^{P} h_p \, e^{j2\pi f_p \frac{n}{N}} \, s[(n - \ell_p)_N] + n[n],
\label{eq:rdisc_int}
\end{equation}
which can be written in matrix form $\mathbf{r} = \mathbf{H}\mathbf{s} + \mathbf{n} \in \mathbb{C}^{N \times 1}$.

The cyclic convolutional matrix representing the doubly dispersive channel is given by
\begin{equation}
\mathbf{H} = \sum_{p=1}^{P} h_p \, \boldsymbol{\Phi}_p\, \mathbf{W}^{f_p} \boldsymbol{\Pi}^{\ell_p} \in \mathbb{C}^{N \times N},
\label{eq:Hmat_full}
\end{equation}
with four constituting parts described in the following.

\vspace{1ex}
\noindent $\blacktriangleright$ $\boldsymbol{\Pi} \in \mathbb{R}^{N \times N}$ is the cyclic forward-shift (delay) matrix,
\begin{equation}
\boldsymbol{\Pi} =
{\renewcommand{\arraystretch}{0.85}
\begin{pmatrix}
0           & \cdots & 0      & 1      \\
1           & \cdots & 0      & 0      \\
\vdots & \ddots & \vdots & \vdots \\
0           & \cdots & 1      & 0
\end{pmatrix}} \in \mathbb{R}^{N \times N},
\label{eq:Pi}
\end{equation}
such that $\mathbf{\Pi}\mathbf{s}$ results in a one-sample cyclic delay of $\mathbf{s}$, and $(\mathbf{\Pi}^{\ell_p})\mathbf{s}$ in a $\ell_p$-sample cyclic delay.

\vspace{1ex}
\noindent$\blacktriangleright$ $\mathbf{W} = \mathrm{diag}(\mathbf{w}) \in \mathbb{C}^{N \times N}$ is the diagonal Doppler phase matrix built from the vector of $N$-th roots of unity
\begin{equation}
\mathbf{w} \triangleq \bigl[1,\; e^{j2\pi/N},\; \ldots,\; e^{j2\pi(N-1)/N}\bigr]^\mathsf{T} \in \mathbb{C}^{N\times 1}.
\label{eq:wvec}
\end{equation}

Since $\mathbf{W}$ is diagonal, its $f_p$-th power equals element-wise exponentiation of $\mathbf{w}$,
\begin{align}
\mathbf{W}^{f_p} &= \mathrm{diag}\!\left(\big[1,\; e^{j2\pi f_p/N},\; \ldots,\; e^{j2\pi f_p(N-1)/N}\big]\right) \in \mathbb{C}^{N \times N}.
\label{eq:Wmat}
\end{align}

\noindent$\blacktriangleright$  $\boldsymbol{\Phi}_p \in \mathbb{C}^{N \times N}$ is a diagonal matrix of \ac{CPP} phases as described in eq.~\eqref{eq:CPP}.
When the prefix does not have a phase correction, as in \ac{OFDM}, $\boldsymbol{\Phi}_p = \mathbf{I}_N$.
For the design choice $2Nc_1 \in \mathbb{Z}$, the \ac{CPP} reduces to a conventional \ac{CP} and $\boldsymbol{\Phi}_p = \mathbf{I}_N$.
For simplicity, and without loss of generality, we consider this design choice met and $\boldsymbol{\Phi}_p = \mathbf{I}_N$ in the following.

\vspace{1ex}
\noindent $\blacktriangleright$ $h_p \in \mathbb{C}$ is the per-path complex fading coefficient.

\subsection{Full Input-Output Circular Convolutional Channel Matrix}
\label{sec:conv_chan}

The full input-output channel matrix (omitting noise) is
\begin{equation}
\mathbf{H} = \sum_{p=1}^{P} h_p \, \mathbf{W}^{f_p} \boldsymbol{\Pi}^{\ell_p} \in \mathbb{C}^{N \times N}.
\label{eq:Hmat}
\end{equation}

\begin{figure*}[b]
    \vspace{-1ex}
\centering
\begin{subfigure}{\textwidth}
\centering
\includegraphics[width=0.695\textwidth]{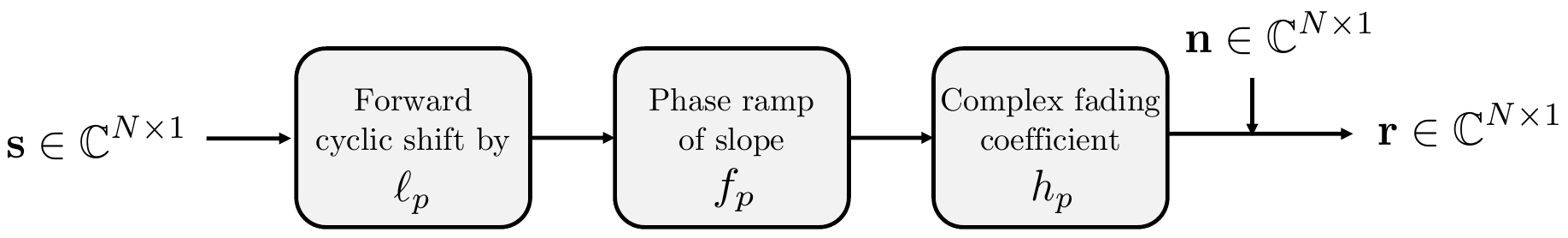}
\caption{Doubly-dispersive channel $\mathbf{H} = \sum_p h_p \mathbf{W}^{f_p}\boldsymbol{\Pi}^{\ell_p}$: paths with same delay collide on the same off-diagonal (\textit{delay collision}).}
\label{fig:convscheme}
\end{subfigure}

\vspace{2ex}
\begin{subfigure}{\textwidth}
\centering
\includegraphics[width=0.785\textwidth]{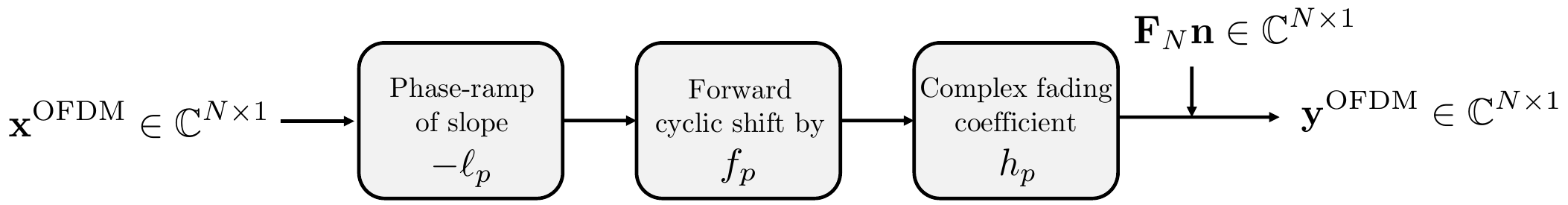}
\caption{\acs{OFDM} effective channel $\mathbf{G}^{\mathrm{OFDM}} = \mathbf{F}_N \mathbf{H} \mathbf{F}_N^\mathsf{H}$: the \ac{IDFT}/\ac{DFT} pair swaps delay and Doppler roles; paths with same Doppler collide on the same off-diagonal (\textit{Doppler collision}).}
\label{fig:ofdmscheme}
\end{subfigure}

\vspace{2ex}
\begin{subfigure}{\textwidth}
\centering
\includegraphics[width=0.865\textwidth]{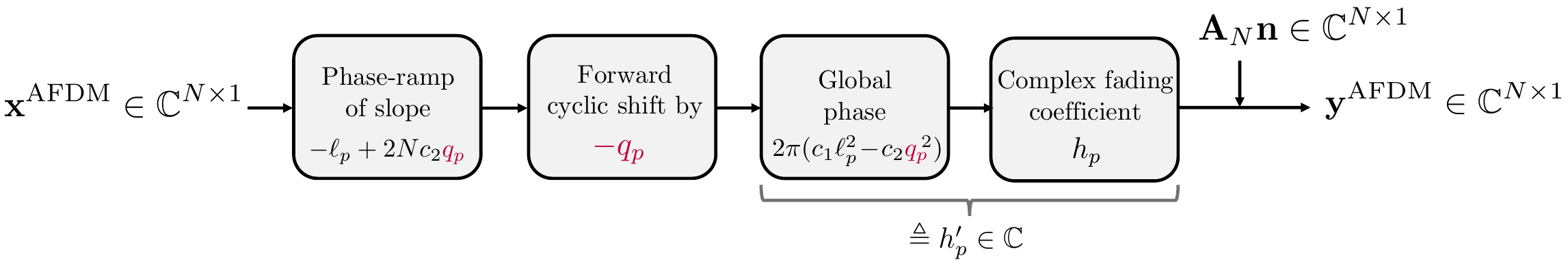}
\caption{\acs{AFDM} effective channel $\mathbf{G}^{\mathrm{AFDM}} = \mathbf{A}_N \mathbf{H} \mathbf{A}_N^\mathsf{H}$: sheared spreading maps each delay-Doppler path onto a unique off-diagonal via $\textcolor{niceMagenta}{q_p = (2f_{\max}+1)\ell_p - f_p}$, eliminating all collisions under the no-aliasing condition.}
\label{fig:afdmscheme}
\end{subfigure}
\caption{Comparison of (a) the doubly-dispersive channel, (b) \acs{OFDM} effective channel, and (c) \acs{AFDM} effective channel effects.}
\label{fig:scheme_compare}
\end{figure*}

As schematically represented in Fig.~\ref{fig:convscheme} for each $p$-th path -- $\boldsymbol{\Pi}^{\ell_p}$ delays the frame by $\ell_p$ samples cyclically, and $\mathbf{W}^{f_p}$ imparts a per-sample phase ramp of slope $f_p/N$, the discrete signature of Doppler.

\subsection{\acs{OFDM} Effective Channel}
\label{sec:ofdm_eff}

In the \ac{OFDM} transceiver, the \ac{IDFT} is applied at the transmitter and the \ac{DFT} at the receiver, yielding $\mathbf{y} = \mathbf{F}_N (\mathbf{H} \mathbf{F}_N^\mathsf{H} \mathbf{x} + \mathbf{n}) \in \mathbb{C}^{N \times 1}$.
The \ac{OFDM} effective channel, omitting noise\footnote{As $\mathbf{F}_N$ is unitary, $\mathbf{F}_N\mathbf{n}$ follows the same distribution as \ac{AWGN}.}, is
\begin{align}
\mathbf{G}^{\mathrm{OFDM}} = \mathbf{F}_N \mathbf{H} \mathbf{F}_N^\mathsf{H}
= \sum_{p=1}^{P} h_p \, \mathbf{F}_N \mathbf{W}^{f_p} \boldsymbol{\Pi}^{\ell_p}
\mathbf{F}_N^\mathsf{H} \in \mathbb{C}^{N \times N}.
\label{eq:Gofdm}
\end{align}

To reveal how delay and Doppler transform under the \ac{DFT}/\ac{IDFT} pair, we expand the per-path contribution ($\forall p$) as
\begin{align}
\!\!\mathbf{H}_p &\triangleq  h_p \, \mathbf{F}_N( \mathbf{W}^{f_p} \boldsymbol{\Pi}^{\ell_p})
\mathbf{F}_N^\mathsf{H} =  h_p \, \mathbf{F}_N (\mathbf{W}^{f_p}  \overbrace{\mathbf{F}_N^\mathsf{H} \mathbf{F}_N}^{\triangleq \, \mathbf{I}_N} \boldsymbol{\Pi}^{\ell_p} )\mathbf{F}_N^\mathsf{H} \nonumber \\[1ex]
&\!\!\!\!\!\!\!\!\!= h_p \big(\mathbf{F}_N \mathbf{W}^{f_p} \mathbf{F}_N^\mathsf{H}\big) \big(\mathbf{F}_N \boldsymbol{\Pi}^{\ell_p} \mathbf{F}_N^\mathsf{H}\big) = h_p \big(\boldsymbol{\Pi}^{f_p}\big) \big( \mathbf{W}^{-\ell_p}\big),\!\!
\label{eq:ofdm_perpath}
\end{align}
where the \ac{DFT} sandwiching swaps the roles of delay and Doppler, as seen by comparing Fig.~\ref{fig:convscheme} and Fig.~\ref{fig:ofdmscheme}: the delay $\ell_p$ now appears as a negative phase ramp, and Doppler $f_p$ affects the cyclic shift.

\textbf{Remark:} The identities $\mathbf{F}_N \mathbf{W}^{f_p} \mathbf{F}_N^\mathsf{H} = \boldsymbol{\Pi}^{f_p}$ and $\mathbf{F}_N \boldsymbol{\Pi}^{\ell_p} \mathbf{F}_N^\mathsf{H} = \mathbf{W}^{-\ell_p}$ hold for integer $f_p$ and $\ell_p$, respectively, following from the \ac{DFT} shift theorem.

The integer delay assumption is reasonable in practice. The integer Doppler assumption, however, is not: for fractional $f_p$, the Doppler energy spreads across adjacent off-diagonals in a Dirichlet-kernel pattern,
\begin{equation}
\bigl(\mathbf{F}_N \mathbf{W}^{f_p} \mathbf{F}_N^\mathsf{H}\bigr)_{k,k'}
= \frac{1}{N}\,\frac{1 - e^{j2\pi(f_p-(k-k'))}}{1 - e^{j2\pi(f_p-(k-k'))/N}},
\label{eq:dirichlet}
\end{equation}
which is centered on the $f_p$-th off-diagonal.

When $f_p = 0$ for all paths, $\mathbf{G}^{\mathrm{OFDM}}$ is diagonal, enabling one-tap equalization.
When $f_p \neq 0$, paths spread onto adjacent subcarriers, causing \acs{ICI}.

More critically, any two paths $p \neq p'$ with $f_p = f_{p'}$ but $\ell_p \neq \ell_{p'}$ resolve to the \emph{same} off-diagonal position of $\mathbf{G}^{\mathrm{OFDM}}$ -- a \emph{Doppler collision}.
This reduces the effective rank of the channel below $P$, preventing full diversity and -- equally importantly for \ac{ISAC} -- preventing unique identification of individual scattering paths, as each off-diagonal conflates multiple targets with the same Doppler shift.
While \ac{OFDM}-based radar sensing has been studied~\cite{SturmProcIEEE2011}, it relies on slow-time, cross-frame processing to extract range-velocity information -- \ac{AFDM}, by contrast, recovers delay-Doppler parameters directly from the single-frame effective channel structure, enabling per-frame \ac{ISAC} without additional processing overhead.
\vspace{-1ex}

\subsection{\acs{AFDM} Effective Channel}
\label{sec:afdm_eff}

In the \ac{AFDM} transceiver, the \ac{IDAFT} $\mathbf{A}_N^\mathsf{H} = \boldsymbol{\Lambda}_{c_1}^\mathsf{H}\mathbf{F}_N^\mathsf{H}\boldsymbol{\Lambda}_{c_2}^\mathsf{H}$ is applied at the transmitter and the \ac{DAFT} $\mathbf{A}_N = \boldsymbol{\Lambda}_{c_2}\mathbf{F}_N\boldsymbol{\Lambda}_{c_1}$ at the receiver, where $\boldsymbol{\Lambda}_{c_i} \triangleq \mathrm{diag}(\boldsymbol{\lambda}_{c_i})$.
The received signal is $\mathbf{y} = \mathbf{A}_N(\mathbf{H}\mathbf{A}_N^\mathsf{H}\mathbf{x} + \mathbf{n})$, and the \ac{AFDM} effective channel is
\begin{align}
\mathbf{G}^{\mathrm{AFDM}} = \mathbf{A}_N \mathbf{H} \mathbf{A}_N^\mathsf{H}
= \sum_{p=1}^{P} h_p \, \mathbf{A}_N \mathbf{W}^{f_p} \boldsymbol{\Pi}^{\ell_p}
\mathbf{A}_N^\mathsf{H} \in \mathbb{C}^{N \times N}. \nonumber \\[-3ex]
\label{eq:Gafdm}
\end{align}

Following the same steps as in \ac{OFDM} in eq.~\eqref{eq:ofdm_perpath} and linear algebra identities (omitted for brevity), the per-path effective channel for \ac{AFDM} is given by

$~$
\vspace{-2ex}
\begin{align}
\!\!\mathbf{H}_p &\triangleq  h_p \, \big(\boldsymbol{\Lambda}_{c_2}\mathbf{F}_N\boldsymbol{\Lambda}_{c_1}\big)( \mathbf{W}^{f_p} \boldsymbol{\Pi}^{\ell_p})\big(\boldsymbol{\Lambda}_{c_1}^\mathsf{H}\mathbf{F}_N^\mathsf{H}\boldsymbol{\Lambda}_{c_2}^\mathsf{H}\big) \\[1ex]
& = \cdots = \underbrace{h_p \, e^{j2\pi (c_1 \ell_p^2 - c_2 q_p^2)}}_{\triangleq \, h'_p} \big(\boldsymbol{\Pi}^{-q_p}\big) \big(\mathbf{W}^{-\ell_p + 2Nc_2 q_p}\big)
\label{eq:afdm_perpath}
\end{align}

\noindent \vspace{-0.5ex} where the newly defined \textit{coupled delay-Doppler} index is
\begin{equation}
q_p \,\triangleq\, 2Nc_1\ell_p - f_p  = (2f_{\max}+1)\ell_p-f_p \in \mathbb{Z},
\label{eq:qp_def}
\end{equation}
with the latter equality following from setting $c_1 = \frac{2f_{\max}+1}{2N}$, the \textit{full diversity} condition of~\cite{Bemani2023TWC}, and $f_{\max}$ being the targeted unambiguous maximum integer Doppler shift.

As illustrated in Fig.~\ref{fig:afdmscheme}, the key difference from \ac{OFDM} is the delay-Doppler index coupling: the off-diagonal position is no longer determined by $f_p$ alone, but by the joint index $q_p$, which encodes both $\ell_p$ and $f_p$.
Paths sharing the same $f_p$ but differing in $\ell_p$ now land on \emph{different} off-diagonals of $\mathbf{G}^{\mathrm{AFDM}}$, resolving the Doppler collision of \ac{OFDM}.

\subsection{Full Diversity Condition and ISAC Implications}
\label{sec:fdc}

Since each path $p$ becomes mapped to a \emph{unique} off-diagonal $-q_p \bmod N$ of $\mathbf{G}^{\mathrm{AFDM}}$, every received sample is a superposition of contributions from \emph{all $P$ paths}, achieving full diversity of order $P$ without any channel coding or precoding~\cite{Bemani2023TWC}.

The \ac{ISAC} implication is direct and equally significant.
Leveraging the unique shift property, one can reverse-engineer the delay-Doppler taps $\ell_p$ and $f_p$ from $q_p$.
Namely, the $P$ peak positions of the off-diagonals of the estimated channel matrix directly yield the radar parameters -- range from delay, velocity from Doppler -- establishing a structural \emph{channel estimation--radar parameter estimation equivalence}~\cite{Rou2024SPMag,Bemani2024WCL,Ni2025TWC}.
This means that \ac{AFDM} provides \ac{ISAC} as an inherent consequence of its waveform structure, not as an add-on.

Crucially, because \ac{AFDM} is a firmware upgrade of \ac{OFDM}, this \ac{ISAC} capability is deployable across the entire installed base of \ac{OFDM} infrastructure, including base stations, Wi-Fi access points, and \ac{V2X} units, without any hardware modification.
This makes \ac{AFDM} a practical path to native \ac{ISAC} at global scale, directly relevant to the IEEE 802.11bf sensing standard and 6G \ac{ISAC} roadmaps~\cite{LiuJSAC2022,ZhangJSTSP2021,Rou2025CommStd}.

Trivially, this diversity result also extends to static (\ac{LTI}) channels, as setting $f_p = 0$ and $c_1 = 1/(2N)$ gives $q_p = \ell_p$, so all static paths land on unique off-diagonals.
\ac{AFDM} thus achieves full multipath diversity of order $P$ in \ac{LTI} channels \emph{without any channel coding} -- a gain that \ac{OFDM} can only approach with rate-reducing channel codes.
This advantage has also been observed in hardware, where software-defined radio deployments in indoor rich-multipath environments show \ac{AFDM} outperforming \ac{OFDM} in uncoded scenarios even in the complete absence of Doppler.

\section{Secondary Waveform Design Space}
\label{sec:secondary}

Inspection of~\eqref{eq:afdm_perpath} reveals that $c_2$ does not appear in the coupling index $q_p$, and therefore does not affect the off-diagonal positions of $\mathbf{G}^{\mathrm{AFDM}}$.
This leaves full diversity and \ac{ISAC} path separability intact for any choice of $c_2$, and makes it a freely available design knob, opening a secondary design space unavailable to \ac{OFDM}. 
To introduce a few for motivation,
\begin{itemize}[noitemsep,topsep=2pt]
\item \textbf{\ac{PAPR} reduction:} per-block pre-chirp selection of $c_2$ minimizes \ac{PAPR} with no impact on diversity or \ac{ISAC}, and the candidate pool grows combinatorially, yielding large \ac{PAPR} gains at low overhead~\cite{Yuan2024PAPR,Ali2025PAPR}.
\item \textbf{Physical-layer security:} $c_2$ randomization or chirp permutation makes eavesdropping exponentially complex, infeasible even for quantum search, and security scales factorially with subcarrier count~\cite{Rou2025PLS,Liu2026PLS}.
\item \textbf{Index modulation:} $c_2$ chosen from a codebook conveys additional bits via the pre-chirp domain at no spectral cost, while preserving subcarrier orthogonality, and index bits have been shown to achieve stronger diversity protection than data symbols~\cite{Liu2024IM,Tao2024JSAC}.
\item \textbf{Noise robustness:} the \ac{DAFT} structure provides resilience to non-white noise and challenging propagation conditions, with \ac{AFDM} consistently outperforming \ac{OFDM} and \ac{OTFS}~\cite{Savaux2025Noise,Cao2024GLOBECOM}.
\item \textbf{Quantum receivers:} \ac{AFDM}'s flexible chirp basis has been shown to enable promising results for Rydberg atomic receivers, enabling quantum-enhanced delay-Doppler estimation which was previously not possible due to the ambiguous delay-Doppler coupling~\cite{Kim2026Rydberg,kim2026multi}.
\end{itemize}

\section{Conclusion}

We highlight that \ac{AFDM} constitutes a software upgrade of \ac{OFDM}, being implementable as a firmware patch over existing infrastructure. It delivers full multipath diversity in both doubly dispersive and static \ac{LTI} channels, native per-frame \ac{ISAC} compatibility through the delay-Doppler path separability of its effective channel, and a broad secondary design space -- including \ac{PAPR} reduction, physical-layer security, and index modulation -- enabled by its free chirp parameters.
These gains are structurally grounded in the \ac{DAFT}-induced delay-Doppler index coupling, which eliminates the path collisions in doubly dispersive multipath environments.
Given the massive installed base of \ac{OFDM} infrastructure worldwide, \ac{AFDM} represents a highly promising, practical path to native \ac{ISAC} at global scale, directly relevant to various standard deployments.

\vspace*{\fill}
\bibliographystyle{IEEEtran}
\bibliography{bib/references}

@article{Bemani2023TWC,
  author    = {A. Bemani and N. Ksairi and M. Kountouris},
  title     = {Affine Frequency Division Multiplexing for Next Generation Wireless Communications},
  journal   = {IEEE Trans. Wireless Commun.},
  volume    = {22},
  number    = {11},
  pages     = {8214--8229},
  month     = nov,
  year      = {2023}
}

@article{Rou2024SPMag,
  author    = {H. S. Rou and others},
  title     = {{From Orthogonal Time-Frequency Space to Affine
               Frequency-Division Multiplexing: A Comparative Study of
               Next-Generation Waveforms for Integrated Sensing and
               Communications in Doubly Dispersive Channels}},
  journal   = {IEEE Signal Process. Mag.},
  volume    = {41},
  number    = {5},
  pages     = {71--86},
  month     = sep,
  year      = {2024}
}

@article{Rou2025CommStd,
  author    = {H. S. Rou and K. R. R. Ranasinghe and V. Savaux and
               G. T. F. de Abreu and D. Gonz{\'a}lez G. and C. Masouros},
  title     = {{AFDM} for {6G}: Properties, Features, and Challenges},
  journal   = {IEEE Commun. Standards Mag.},
  note      = {Early Access},
  doi       = {10.1109/MCOMSTD.2025.3643183},
  year      = {2026}
}

@article{Bemani2024WCL,
  author    = {A. Bemani and N. Ksairi and M. Kountouris},
  title     = {Integrated Sensing and Communications With Affine Frequency Division Multiplexing},
  journal   = {IEEE Wireless Commun. Lett.},
  volume    = {13},
  number    = {5},
  pages     = {1255--1259},
  month     = may,
  year      = {2024},
  doi       = {10.1109/LWC.2024.3367178}
}

@article{Ni2025TWC,
  author    = {Y. Ni and P. Yuan and Q. Huang and F. Liu and Z. Wang},
  title     = {An Integrated Sensing and Communications System Based on
               Affine Frequency Division Multiplexing},
  journal   = {IEEE Trans. Wireless Commun.},
  volume    = {24},
  number    = {5},
  pages     = {3763--3779},
  month     = may,
  year      = {2025},
  doi       = {10.1109/TWC.2025.3532993}
}

@article{LiuJSAC2022,
  author    = {F. Liu and others},
  title     = {Integrated Sensing and Communications: Toward Dual-Functional
               Wireless Networks for {6G} and Beyond},
  journal   = {IEEE J. Sel. Areas Commun.},
  volume    = {40},
  number    = {6},
  pages     = {1728--1767},
  month     = jun,
  year      = {2022},
  doi       = {10.1109/JSAC.2022.3156632}
}

@article{ZhangJSTSP2021,
  author    = {J. A. Zhang and others},
  title     = {An Overview of Signal Processing Techniques for Joint
               Communication and Radar Sensing},
  journal   = {IEEE J. Sel. Topics Signal Process.},
  volume    = {15},
  number    = {6},
  pages     = {1295--1315},
  month     = nov,
  year      = {2021},
  doi       = {10.1109/JSTSP.2021.3113120}
}

@inproceedings{Bemani2022ICASSP,
  author    = {A. Bemani and N. Ksairi and M. Kountouris},
  title     = {Low Complexity Equalization for {AFDM} in Doubly Dispersive Channels},
  booktitle = {Proc. IEEE Int. Conf. Acoustics, Speech Signal Process. (ICASSP)},
  month     = may,
  year      = {2022},
  doi       = {10.1109/ICASSP43922.2022.9746610}
}

@article{Tao2024JSAC,
  author    = {Y. Tao and others},
  title     = {Affine Frequency Division Multiplexing with Index Modulation:
               Full Diversity Condition, Performance Analysis, and
               Low-Complexity Detection},
  journal   = {IEEE J. Sel. Areas Commun.},
  volume    = {43},
  number    = {4},
  pages     = {1041--1055},
  month     = apr,
  year      = {2025},
  doi       = {10.1109/JSAC.2025.3531561}
}

@article{SturmProcIEEE2011,
  author    = {C. Sturm and W. Wiesbeck},
  title     = {Waveform Design and Signal Processing Aspects for Fusion
               of Wireless Communications and Radar Sensing},
  journal   = {Proc. IEEE},
  volume    = {99},
  number    = {7},
  pages     = {1236--1259},
  month     = jul,
  year      = {2011},
  doi       = {10.1109/JPROC.2011.2131110}
}

@article{Rou2026Spreading,
  author    = {H. S. Rou and G. T. F. de Abreu and
               E. Bj{\"o}rnson and S. Kim and M. Kountouris},
  title     = {The Resurrection of Spectrum Spreading for {6G} and Beyond: From Sinusoids to Chirps},
  journal   = {arXiv:2605.00249},
  year      = {2026}
}

@article{Rou2026OpenJ,
  author    = {H. S. Rou and V. Savaux and Z. Sui and
               G. T. F. de Abreu and Z. Liu},
  title     = {{AFDM}: Evolving {OFDM} Towards {6G+}},
  journal   = {Early Access, IEEE Open Journ. Commun. Soc.},
  month     = Sep,
  year      = {2026}
}

@article{Yuan2024PAPR,
  author    = {H. Yuan and others},
  title     = {{PAPR} Reduction With Pre-Chirp Selection for Affine
               Frequency Division Multiplexing},
  journal   = {IEEE Wireless Commun. Lett.},
  note      = {Early Access},
  doi       = {10.1109/LWC.2024.10812762},
  year      = {2025}
}

@article{Ali2025PAPR,
  author    = {A. Ali and A. Arous and H. Arslan},
  title     = {Spreading the Wave: Low-Complexity {PAPR} Reduction for
               {AFDM} and {OCDM} in {6G} Networks},
  journal={IEEE Trans. Green Commun. Net.},
  volume={10},
  pages={1565-1577},
  year      = {2026}
}

@article{Rou2025PLS,
  author    = {H. S. Rou and G. T. F. de Abreu},
  journal={IEEE Wireless Commun. Let.}, 
  title={Chirp-Permuted {AFDM} for Quantum-Resilient Physical-Layer Secure Communications}, 
  year={2025},
  volume={14},
  number={8},
  pages={2376-2380}
}

@article{Liu2026PLS,
  author    = {H. Liu and others},
  title     = {Making {AFDM} Secure Against Eavesdroppers: A Phase Function
               Design Approach},
  journal   = {arXiv:2605.14837},
  year      = {2026}
}

@INPROCEEDINGS{Liu2024IM,
  author={G. Liu and T. Mao and R. Liu and Z. Xiao},
  booktitle={2024 International Wireless Communications and Mobile Computing (IWCMC)}, 
  title={Pre-Chirp-Domain Index Modulation for Affine Frequency Division Multiplexing}, 
  year={2024},
  volume={},
  number={},
  pages={0473-0478}}

@INPROCEEDINGS{Savaux2025Noise,
  author = {V. Savaux and others},
  booktitle={2025 4th International Conference on 6G Networking (6GNet)}, 
  title={On the Noise Robustness of Affine Frequency Division Multiplexing: Analysis and Applications}, 
  year={2025},
  volume={},
  number={},
  pages={65-72}}

@inproceedings{Cao2024GLOBECOM,
  author    = {R. Cao and others},
  title     = {{AFDM} Channel Estimation in Multi-Scale Multi-Lag Channels},
  booktitle = {Proc. IEEE Global Commun. Conf. (GLOBECOM)},
  month     = dec,
  year      = {2024},
  doi       = {10.1109/GLOBECOM52923.2024.10901055}
}

@article{Kim2026Rydberg,
  author    = {H. Kim and H. S. Rou and K. Min and
               G. T. F. de Abreu and S. Kim},
  title     = {Dual-Chirp {AFDM} for Joint Delay-Doppler Estimation with
               {Rydberg} Atomic Quantum Receivers},
  journal   = {2026 IEEE Int. Symp. Personal, Indoor, Mobile Radio Commun. (PIMRC)},
  year      = {2026}
}

@article{Mirabella2026JSAC,
  author    = {M. Mirabella and H. S. Rou and P. Di Viesti and
               G. T. F. de Abreu and G. M. Vitetta},
  title     = {Continuous-Time Analysis of {AFDM}: Fundamental Bounds,
               and Effects of Pulse-Shaping and Hardware Impairments},
  journal   = {arXiv:2602.20909},
  year      = {2026}
}

@article{kim2026multi,
  title={Multi-Chirp {AFDM} for Rydberg Atomic Quantum Receivers: Waveform and Algorithm Design},
  author={H. Kim and H. S. Rou and K. H. Min and G. T. F. de Abreu and S. Kim},
  journal={arXiv:2607.27903},
  year={2026}
}

\end{document}